\documentclass[manuscript,screen,sigplan,nonacm]{acmart}

\usepackage{amsmath}
\usepackage{mathpartir}
\usepackage{amsthm}
\usepackage{soul}
\usepackage{stmaryrd}
\usepackage{hyperref}
\usepackage{cleveref}

\usepackage{paper_alcides}
\usepackage{fbox}
\usepackage{mdframed}

\newcommand{\semanticsargs}[1][]{#1\end{mathpar}}
\newcommand{\semantics}[3][]{\begin{mathpar}\inferrule[#1]{#2}{#3}\semanticsargs}

\theoremstyle{definition}

\newcommand{\ttt}[1]{\texttt{#1}}
\newcommand{\semi}{\ttt{;}}
\newcommand{\knew}{\ttt{new}~}
\newcommand{\kassign}{\mathrel{\ttt{=}}}
\newcommand{\kclass}{\ttt{class}~}
\newcommand{\kextends}{~\ttt{extends}~}
\newcommand{\ksuper}{\ttt{super}}
\newcommand{\kthis}{\ttt{this}}
\newcommand{\kif}{\ttt{if}}
\newcommand{\kelse}{\ttt{else}}
\newcommand{\m}[1]{\overline{#1}}
\newcommand{\set}[1]{\{~ #1 ~\}}
\newcommand{\block}[1]{\{~ #1 ~\}}
\newcommand{\kvoid}{\ttt{void}}

\newcommand{\knull}{\ttt{null}}

\newcommand{\kreturn}{\ttt{return}~}

\newcommand{\keq}{\mathrel{\ttt{==}}}

\newcommand{\fields}{\mathsf{fields}}
\newcommand{\ftype}{\mathsf{ftype}}

\newcommand{\mtype}{\mathsf{mtype}}

\newcommand{\owned}{\mathsf{owned}}
\newcommand{\alias}{\mathsf{alias}}
\newcommand{\unique}{\mathsf{unique}}
\newcommand{\aliasfn}{\mathrm{alias}}
\newcommand{\shared}{\mathsf{shared}}

\DeclareMathOperator{\unify}{unify}

\DeclareMathOperator{\dom}{dom}

\newcommand{\own}[1]{\owned\,\ttt{#1}}
\newcommand{\uniq}[1]{\unique\,\ttt{#1}}
\newcommand{\ali}[2]{\alias({\scriptsize \ttt{#1}})\,\ttt{#2}}
\newcommand{\annot}[1]{\hspace{1em}\colorbox{pink!30}{\footnotesize #1}}

\let\origthelstnumber\thelstnumber
\makeatletter
\newcommand*\Suppressnumber{%
  \lst@AddToHook{OnNewLine}{%
    \let\thelstnumber\relax%
     \advance\c@lstnumber-\@ne\relax%
    }%
}

\newcommand*\Reactivatenumber{%
  \lst@AddToHook{OnNewLine}{%
   \let\thelstnumber\origthelstnumber%
   \setcounter{lstnumber}{\numexpr\thelstnumber+1\relax}%
  }%
}

\AtBeginDocument{%
  \providecommand\BibTeX{{%
    \normalfont B\kern-0.5em{\scshape i\kern-0.25em b}\kern-0.8em\TeX}}}

\setcopyright{acmcopyright}
\copyrightyear{2018}
\acmYear{2018}
\acmDOI{XXXXXXX.XXXXXXX}

\acmConference[Conference acronym 'XX]{Make sure to enter the correct
  conference title from your rights confirmation emai}{June 03--05,
  2018}{Woodstock, NY}
\acmBooktitle{Woodstock '18: ACM Symposium on Neural Gaze Detection,
 June 03--05, 2018, Woodstock, NY} 
\acmPrice{15.00}
\acmISBN{978-1-4503-XXXX-X/18/06}

\begin{document}

\title{Latte: Lightweight Aliasing Tracking for Java}


\author{Conrad Zimmerman}
\authornote{Both authors contributed equally to this research.}
\affiliation{%
  \institution{Brown University}
  \streetaddress{TODO}
  \city{Providence}
  \state{RI}
  \country{USA}}
\email{conrad_zimmerman@brown.edu}

\author{Catarina Gamboa}
\authornotemark[1]
\affiliation{%
  \institution{Carnegie Mellon University, and\\Faculdade de Ciências da Universidade de Lisboa}
  \city{Pittsburgh}
  \state{PA}
  \country{USA}
}\email{cgamboa@andrew.cmu.edu}

\author{Alcides Fonseca}
\affiliation{%
 \institution{Faculdade de Ciências da Universidade de Lisboa}
 \city{Lisbon}
 \country{Portugal}
 } \email{amfonseca@fc.ul.pt}

\author{Jonathan Aldrich}
\affiliation{%
  \institution{Carnegie Mellon University}
  \city{Pittsburgh}
  \state{PA}
  \country{USA}
  }  \email{jonathan.aldrich@cs.cmu.edu}


\renewcommand{\shortauthors}{}

\begin{abstract}
Many existing systems track aliasing and uniqueness, each with their own trade-off between expressiveness and developer effort.

We propose Latte,
 a new approach that aims to minimize both the amount of annotations and the complexity of invariants necessary for reasoning about aliasing in an object-oriented language with mutation. Our approach only requires annotations for parameters and fields, while annotations for local variables are inferred.
Furthermore, it relaxes uniqueness to allow aliasing among local variables, as long as this aliasing can be precisely determined. This enables support for destructive reads without changes to the language or its run-time semantics.

Despite this simplicity, we show how this design can still be used for tracking uniqueness and aliasing in a local sequential setting, with practical applications, such as modeling a stack.

\end{abstract}

\ccsdesc[500]{Theory of computation~Program semantics}
\ccsdesc[500]{Software and its engineering~Object oriented languages}

\keywords{aliasing, uniqueness, ownership, java}


\received{20 February 2007}
\received[revised]{12 March 2009}
\received[accepted]{5 June 2009}

\maketitle

\section{Introduction}


From low-level languages like C to high-level programming languages like Python, the combination of mutability with aliasing has been the source of many bugs, warranting its own term (\href{https://martinfowler.com/bliki/AliasingBug.html}{Aliasing Bug}).


Reasoning about aliasing is difficult, as it usually requires a global analysis of the program and its possible traces of execution. 
To overcome this challenge, the community has proposed type systems that track and restrict aliasing~\cite{geneva_convention,DBLP:series/lncs/7850}. 
In this very large design space, there are lines of work more focused on uniqueness, ownership and permissions. 
However, these proposals add complexity relative to ordinary type systems, and in some cases require developers to understand and reason about quite complex concepts.
For example, ownership has been frequently mentioned as one of the hardest concepts to learn in the Rust programming language~\cite{Klabnik_Nichols_23, blog_learn_rust}.

Thus, we propose a type system for uniqueness and aliasing that aims to be more usable and impose low overhead on developers.
Moreover, we intend to keep our approach as simple as possible so that it can support the development of more complex type systems (like Liquid Types~\cite{DBLP:conf/pldi/RondonKJ08}) that require reasoning about uniqueness and aliasing.

In particular, we propose a type system for a subset of the Java language that tracks uniqueness and heap aliasing with low annotation effort and no necessary runtime changes. 
To handle uniqueness, we provide the developer with the simple invariant that \textit{no two references in the heap point to the same unique object}. 

We require few annotations, specifically two ($\unique$ and $\shared$) for object fields and return types, and three (adding $\owned$) for method parameters, and we infer the remaining information for local variables.

Moreover, we aim to maintain Java's runtime semantics unaltered, and therefore do not build in destructive reads; the programmer can, however, get a similar effect with an explicit assignment to \ttt{null}.

In the remainder of the paper, we present an overview of previous work that is closely related to ours (\Cref{sec:related-work}), followed by our approach with the presentation of an example and the system's grammar and typing rules (\Cref{sec:approach}). At the end, we discuss some of the system's limitations and future directions for this work (\Cref{sec:futurework}).

\section{Related Work}\label{sec:related-work}
There are different approaches for managing aliasing in programming languages.  
One popular line of work focuses on ownership types~\cite{DBLP:conf/oopsla/ClarkePN98}, a type system that restricts access to objects according to their owners. 
There have been multiple proposals for ownership types with different flavors~\cite{DBLP:conf/ecoop/ClarkeW03}.  Unfortunately, classic ownership types alone do not track aliasing within ownership boundaries, making it difficult for verification tools to precisely track the effect of assignments.

Tracking uniqueness can provide strong guarantees about aliasing, which is useful for verification as
well as safe manual memory management, e.g. as implemented in Rust~\cite{rust}.
However, more powerful systems such as Rust's (which is called ``ownership,'' though it provides uniqueness
in the sense described in the research literature) are known to be
complex and difficult for developers to understand~\cite{usability_ownership}. 

Uniqueness types \cite{DBLP:conf/ifl/VriesPA07} allow uniqueness properties to be specified as part of a data type. However, this approach is tailored to functional programming languages and requires a significantly different type system, thus its applicability to Java is limited.

Other type systems for uniqueness focus on the use cases of concurrency and message passing.
For example, Haller and Odersky~\cite{DBLP:conf/ecoop/HallerO10} use capabilities~\cite{DBLP:conf/ecoop/BoylandNR01} to add uniqueness and borrowing to a Java-like languages with a focus on message passing in concurrent object-oriented programming. They use a concept of \emph{separate uniqueness}, where distinct variables do not share a common reachable object. Thus uniqueness is used to enforce separation, which is desirable for concurrent message-passing systems.

More recently, \citet{DBLP:conf/pldi/MilanoTM22} presented a new language and type system for safe concurrency by statically ensuring that different threads cannot access the same heap regions. Their proposal focused on reducing the annotation burden and eliminating the need for unnatural rewrites required by more restrictive programming models.
Also aiming for a minimal set of annotations, LaCasa \cite{DBLP:conf/oopsla/HallerL16} adds uniqueness to the Scala language using object capabilities. However, this approach requires classes to adhere to the object-capability discipline, and their empirical evaluation showed that most classes from the standard library do not follow these rules.

There are other approaches that focus on modeling different aspects of aliasing. 
Reachability types~\cite{DBLP:journals/pacmpl/BaoWBJHR21} uses \emph{reachability sets} to reason about ownership, and tracks reachable values using type qualifiers. This work layers uniqueness, nested mutable state and other concepts over the tracking of reachability sets. Unique accesses are enabled by killing all other access paths to a reference.

\citet{DBLP:conf/ecoop/CastegrenW16} combined many of the previously-mentioned concepts in their $\kappa$ language. This language uses reference capabilities to ensure separation, and combines techniques from ownership types, linear types~\cite{DBLP:conf/ifip2/Wadler90}, and regions~\cite{DBLP:conf/ecoop/GreenhouseB99} in a concurrent and parallel object-oriented setting.

The systems described above were not implemented for Java, however, and it is unclear how to do so, as they rely on language features that Java does not have, such as capabilities in Scala or a primitive swap operation.

AliasJava~\cite{DBLP:conf/oopsla/AldrichKC02} does extend Java with type annotations in Java that specify data sharing relationships. The type system includes four annotations: \textit{unique}, \textit{owned}, \textit{lent} and \textit{shared}, and reduces the annotation burden by inferring annotations.  Our approach is similar in spirit, but achieves greater simplicity by doing without ownership and ownership parameters, while allowing more local alising within a method.

Many early systems for uniqueness used destructive reads, but these often negatively impact program complexity by reducing the ability to query information contained in objects \cite{DBLP:conf/popl/BoylandR05}.
Therefore, alias burying~\cite{DBLP:journals/spe/Boyland01} aims to define a uniqueness system for Java-like languages without using destructive reads by relying on the idea that aliases that will not be used again can be buried. However, as noted by \citet{DBLP:conf/popl/BoylandR05}, the analysis described in the initial work~\cite{DBLP:journals/spe/Boyland01} exposes implementation details, such as the fields read by a method, which breaks encapsulation and modularity.

In summary, the prior work has one of three limitations: reliance on language or type system features not present in Java, modularity or coding pattern issues, or a larger and more complex set of abstractions for programmers to understand compared to our goals.  All of these design choices raise the adoption cost for developers.
In our approach, Latte, we try to address these difficulties by creating a lightweight uniqueness system with few annotations. 
Latte, which we present in the following section, requires no changes to the language semantics, and allows many common code patterns while precisely tracking aliasing.


\section{Approach}\label{sec:approach}

As we described previously, our design aims to impose minimal restrictions while enforcing unique references (in the heap) and tracking aliasing (in the stack). While our system is not as expressive as others in previous work, its main advantages are an easily-understood programming model and low annotation complexity.

In particular, our design only requires annotations on fields and parameters (with only two and three possible choices, respectively). This burden can be further reduced by choosing sensible defaults. Local variables do not need annotations, as the aliasing between local variables and field values is inferred, which reduces the barrier to adopting this system.

In this section, we first give a high-level description of our approach, and then use an example to build intuition about our model. We then formally define the typing rules on top of a Featherweight Java~\cite{DBLP:journals/toplas/IgarashiPW01}-inspired core language and explain how these rules result in the intended behavior. Finally, we demonstrate the expressive power of our system with a more complex example.

\subsection{Description}\label{sec:description}

First, we restrict our definition of uniqueness to only consider reachable values on the heap, thus unique values may be stored in at most one reachable heap location, and aliased in the local environment. However, these \emph{dynamic aliases} \cite{DBLP:conf/oopsla/Hogg91} may only be used as long as such aliasing can be precisely inferred. Our treatment of dynamic aliases and unreachable heap locations is similar to that of alias burying \cite{DBLP:conf/popl/BoylandR05}.

In Latte, the annotation $\unique$ is used to identify unique values (as defined above). $\owned$ identifies borrowed values, since it is only used on method parameters whose value will be owned by some other context when entering the method body. $\shared$ identifies values that may or may not be unique.

We aim to use our analysis in an automated verifier such as LiquidJava~\cite{liquidjava} to reason about mutation of unique (and borrowed) values. This requires precisely identifying all values that may be affected by a particular mutation. Our approach does this while permitting dynamic aliases, by inferring annotations of the form $\alias(p)$ or $\bot$ during type checking. These special annotations are only inferred; they are never written by the developer.

For each local variable $x$, our typing environment $\Delta$ contains a class $C$ and an annotation $\alpha$ which describes the uniqueness of $x$ at that point. The formal definition of $\Delta$ is given later in \Cref{fig:typing-environment}.

\subsection{Example}

\begin{figure}
\begin{mdframed}[innertopmargin=7pt,innerleftmargin=15pt,innerrightmargin=0pt,innerbottommargin=7pt]
{\footnotesize
\begin{lstlisting}[aboveskip=0pt,belowskip=0pt]
class Node {
  unique Object value;
  unique Node next;

  Node(unique Object value, unique Node next)
  {  this.value = value; 
     this.next = next; }
}

class Stack {
  unique Node root;

  Stack(unique Node root)
  { this.root = root; }

  void push(owned Stack this,
                unique Object value) {<@\label{ln:push-start}\Suppressnumber@>
    <@\annot{$\Delta = \ttt{this} : \own{Stack}, ~ \ttt{value} : \uniq{Object}$} \Reactivatenumber@>
    Node r; Node n; <@\label{ln:push-decl}\Suppressnumber@>
    <@\annot{$\Delta = \cdots, ~\ttt{r} : \bot\,\ttt{Node},~ \ttt{n} : \bot\,\ttt{Node}$} \Reactivatenumber@>
    r = this.root; <@\label{ln:r-init}\Suppressnumber@>
    <@\annot{$\Delta = \cdots,~ \ttt{r} : \ali{this.root}{Node}$} \Reactivatenumber@>
    this.root = null; <@\label{ln:root-clear}\Suppressnumber@>
    <@\annot{$\Delta = \cdots,~ \ttt{r} : \uniq{Node}$}\Reactivatenumber@>
    n = new Node(value, r); <@\label{ln:n-init}\Suppressnumber@>
    <@\annot{$\begin{aligned}
        \Delta =\,& \ttt{this} : \own{Stack},~
          \ttt{value} : \bot\,\ttt{Object}, \\
          &\ttt{r} : \bot\,\ttt{Node},~
          \ttt{n} : \uniq{Node}
    \end{aligned}$}\Reactivatenumber@>
    this.root = n;<@\label{ln:root-assign}\Suppressnumber@>
    <@\annot{$\Delta = \cdots, \ttt{n} : \ali{this.root}{Node}$}\Reactivatenumber@>
  }

  unique Object pop(owned Stack this) {
    Object value;
    if (this.root == null) {
        value = null;
    } else {
        value = this.root.value; <@\label{ln:pop-value-assign}@>
        Node next;
        next = this.root.next; <@\label{ln:pop-next-assign}@>
        this.root = next; <@\label{ln:pop-next-update}@>
    } <@\label{ln:pop-if-exit}@>
    return value;
  }
}
\end{lstlisting}}
\end{mdframed}
\caption{Example: a stack for unique references}
\label{fig:stack-example}
\end{figure}

We illustrate the main features of our approach by implementing push and pop operations for a stack storing unique values. References to objects pushed onto the stack may not be stored on the heap anywhere else. This invariant could be used by an automated verifier to show that values pushed onto the stack will not be mutated until they are returned by pop. The code is shown in \Cref{fig:stack-example}.

First, we demonstrate how the \ttt{push} method is validated. At each step, we have listed (in \Cref{fig:stack-example}) the current typing context $\Delta$ to illustrate the verification process.

The typing environment at the beginning of the method body contains
the parameters and their types (line \ref{ln:push-start}). Because the two variables \ttt{r} and \ttt{n} are declared, but not yet initialized at line \ref{ln:push-decl}, they are annotated with $\bot$ to mark them inaccessible. \ttt{r} is aliased with \ttt{this.root} at line \ref{ln:r-init}, and this aliasing information is added to the environment by annotating \ttt{r} with $\alias(\ttt{this.root})$. Next, \ttt{this.root} is assigned, which \emph{isolates} it in the typing environment (a process that we describe in \S\ref{sec:isolation}). This invalidates all aliases to it, which allows the previously-aliased variable \ttt{r}
to become $\unique$ and thus claim ownership.

Note that lines \ref{ln:r-init}-\ref{ln:root-clear} are equivalent to a destructive read. However, we do not need to change the language semantics or introduce new language constructs. Also, the need for this destructive read is easily understood: we want to store \ttt{this.root} in a different place in the heap, thus we need to remove its current value from the heap. Otherwise, our uniqueness invariant would be violated since the value at \ttt{this.root} (which is declared $\unique$) would be stored in multiple reachable places in the heap.

Continuing with our example, we initialize a new \ttt{Node} object at line \ref{ln:n-init}. The constructor of \ttt{Node} has the signature $(\unique~\ttt{Object}, \unique~\ttt{Node})$, which states that it \emph{consumes} both arguments. 
This marks \ttt{value} and \ttt{r} as inaccessible ($\bot$) in the calling context. We encapsulate constructor bodies, thus we do not know what \ttt{value} or \ttt{r} may be aliased with after they are passed to the constructor. Since we cannot track this aliasing, any usage of these values is disallowed after they are passed to the constructor.

Finally, we assign \ttt{n} to \ttt{this.root} at line \ref{ln:root-assign}. Since our typing context only stores annotations for local variables, and not for fields, we update the annotation for the local variable \ttt{n}, which is on the RHS of the assignment. The annotation $\alias(\ttt{this.root})$ simply denotes that its target \ttt{this.root} contains the same value as the annotated variable \ttt{n}, thus it does not matter which side of an alias is annotated.
After this line, $\Delta$ indicates that \ttt{n} is aliased with \ttt{this.root}, thus we have precisely determined all local aliases to the $\unique$ value \ttt{this.root}, and ensured that \ttt{this.root} is stored at only one location on the heap.

This first method gives an overview of our approach; the second method (\lstinline|pop|, at line \ref{ln:pop-next-assign}) will be presented in \Cref{sub-sec:statement-typing} after the grammar and typing rules are introduced.


\subsection{Grammar}
\begin{figure}
\begin{mdframed}[innertopmargin=0pt]
\begin{align*}
    P        ~::=&~ \m{CL} \\
    CL        ~::=&~ \kclass C \kextends C ~ \block{ \m{F} ~ K ~ \m{M} } \\
    \alpha_f ~::=&~ \unique \mid \shared \\
    \alpha_p ~::=&~ \unique \mid \shared \mid \owned \\
    F        ~::=&~ \alpha_f ~C~ f\semi \\
    K        ~::=&~ C(\m{\alpha_f~C~g}, \m{\alpha_f ~ C ~ f})
                 \block{ \ksuper(\m{g})\semi \m{\kthis.f = f\semi} } \\
    \tau     ~::=&~ \alpha_f ~ C \mid \kvoid \\
    M        ~::=&~ \kvoid ~ m(\alpha_p~C~\kthis, ~ \m{\alpha_p ~ C ~ x}) ~ \block{ \m{s} } \\
             \mid&~ \alpha_f ~ C ~ m(\alpha_p~C~\kthis, ~ \m{\alpha_p ~ C ~ x}) ~ \block{ \m{s} ~ \kreturn~e\semi } \\
    p        ~::=&~ x \mid p.f \\
    e        ~::=&~ \knull \mid p \\
    s        ~::=&~ C ~ x\semi
             \mid~ x \kassign e\semi
             \mid   x.f \kassign e\semi \\
             \mid&~ x \kassign \knew C(\m{e})\semi
             \mid x \kassign x.m(\m{e})\semi
             \mid x.m(\m{e})\semi \\
             \mid&~ \kif~(e \keq e)~ s ~\kelse~ s
             \mid \block{ \m{s} } 
  \end{align*}
\end{mdframed}
\caption{Grammar extended from Featherweight Java}\label{fig:grammar}
\end{figure}

For our grammar, presented in \Cref{fig:grammar}, we extended Featherweight Java~\cite{DBLP:journals/toplas/IgarashiPW01} with statements, including field and variable assignments as in Java, to better approximate the Java language in terms of mutability (a key concern in this paper). To model our particular system, we added the $\unique$, $\shared$, and $\owned$ annotations. All fields ($F$) must be annotated with either $\unique$ or $\shared$ ($\alpha_f$), while method parameters (in $M$) must be annotated with one of the three annotations ($\alpha_p$). Note that variable declarations are not annotated (first production of $s$).


Our $\owned$ annotation is often called \textit{lent} or \textit{borrowed} in other systems. Our choice reflects the state of the value within the method body -- the value is owned by some other context. 

\subsection{Typing rules}
Our typing rules use a local type environment $\Delta$. 
This environment maps variables to an annotated class ($ \alpha~C$), where the annotation specifies the current aliasing or uniqueness information. 
The form of $\Delta$ is given in \Cref{fig:typing-environment}. 

A $\shared$ annotation denotes that the variable can be accessed by outside objects -- untracked aliases may exist. $\owned$ denotes that the value of the variable is borrowed; specifically, its value is unique in the current context and no new aliases may be added to the heap. $\unique$ denotes ownership -- the value is only stored at this location (modulo precisely-tracked dynamic aliases). A local variable annotated $\unique$ may be converted to a $\shared$, based on its usage. $\bot$ denotes that the value is inaccessible.

\begin{figure}[h]
    \begin{mdframed}[innertopmargin=0pt]
        \begin{align*}
            \alpha_e &::= \owned \mid \shared \mid \unique \mid \unique(p.f) \\
            \alpha &::= \owned \mid \shared \mid \unique \mid \alias(p) \mid \bot \\
            \Delta &::= \cdot \mid x : \alpha ~ C, ~ \Delta
        \end{align*}
    \end{mdframed}
    \caption{Typing environment and annotations used in typing rules}
    \label{fig:typing-environment}
\end{figure}

\subsubsection{Aliasing}\label{sec:aliasing}

Aliasing between variables and fields is tracked by entries of the form $x : \alias(p)$ in $\Delta$. This denotes that the $x$ stores the same value as the \emph{path} (a variable or some field access) $p$. Two paths are aliased iff they reference the same object, thus aliasing is an equivalence relation. This is encoded by the judgment $\Delta \vdash p_1 \equiv p_2$, which denotes that $\Delta$ indicates that the path $p_1$ is aliased with $p_2$. Formal rules are given in \Cref{sec:aliasing-rules}.


Given an environment $\Delta$, we define its \textit{alias graph} to be a (undirected) graph whose nodes are syntactic paths ($p$ as defined in \Cref{fig:grammar}), and distinct paths $p_1$ and $p_2$ are connected iff $\Delta \vdash p_1 \equiv p_2$. Each component of this graph may contain at most one path annotated with $\owned$, $\unique$, or $\shared$.


Intuitively, the alias graphs for each program point (which identify allowable aliasing), along with validation of uniqueness invariants (which ensures that no other aliases of unique values exists), is the primary product of our analysis. This output can then be used to automatically verify the effects of mutation or, more generally, separation invariants.

\subsubsection{Side note: concurrency}\label{sec:concurrency}
Since we are tracking aliases across multiple statements, and our alias annotations may point to mutable heap locations, it may seem challenging to handle concurrency. However, we only claim to precisely track aliases of unique or borrowed values, i.e. expressions for which $\Delta \vdash e : \owned~C \dashv \Delta'$ holds for some $C$ and $\Delta'$. Intuitively, if this holds for a variable $x$, and $x$ is aliased to $y.f$, then $y.f$ is also unique, which requires $y$ to be unique. In other words, either the current context or some calling context is the sole owner of $y$, and thus of the heap location $y.f$. (This reasoning may be extended for $y.f.g$, etc.) Therefore we can determine all mutations that would affect this alias relation. In other words, assuming soundness of our approach for sequential programs, it should remain sound for concurrent programs, as long as unique values accessible to the spawned thread are consumed after a fork operation.

\subsubsection{Reachable aliasing}

$\Delta \vdash p_1 \approxeq p_2$ denotes that a value reachable from $p_1$ (for example, $p_1.f$) may be aliased with a value reachable from $p_2$ (for example, $p_2.f.g$). Formal rules are given in appendix \ref{sec:reachable-aliasing-rules}. If $\Delta \nvdash p_1 \approxeq p_2$ then $p_1$ and $p_2$ are \textit{separately unique}, as defined in \cite{DBLP:conf/ecoop/HallerO10}.

\subsubsection{Expression typing}\label{sec:expr-typing}

$\Delta \vdash e : \alpha_e ~C \dashv \Delta'$ denotes that $e$ may be used as a value with class $C$ and ownership annotation $\alpha_e$, provided that all future typing uses the $\Delta'$ typing environment. Formal rules are given in \Cref{fig:expression-usage-rules}.

$\Delta \vdash e : \unique(p.f) \dashv \Delta'$ denotes that $e$ refers to a unique value (as defined in \S\ref{sec:description}), and $e$ is aliased with $p.f$ in $\Delta'$.
This is used to validate assignments to $\unique$ fields, since the assignee is a field whose annotation is not stored in $\Delta$. Thus aliasing information is tracked by annotating the assignment value, instead of annotating the assignee.

For a variable $x$ annotated with $\alias(p)$, $\Delta \vdash x : \alpha_e~C \dashv \Delta'$ holds if and only if $\Delta \vdash p : \alpha_e~C \dashv \Delta'$ -- in other words, aliased variables may be used exactly how the path they alias may be used.

When borrowing a $\unique$ field value (i.e. passing the value to a parameter annotated as $\owned$), the object reference must also be $\unique$.
For example, if we have variables $x : \shared~C$ and $y : \shared~C$, and $C$ contains a $\unique$ field $f$, we cannot borrow $x.f$ because we do not know whether the same heap location is already borrowed through $y.f$. Thus one can introduce aliases to a $\unique$ field of a $\shared$ value, but those values can only be used after a destructive read or some equivalent operation.

Finally, a value of a subtype $C$ may be used as a value of the supertype type $D$ with the same annotation.

\begin{figure}[h]
    \begin{mdframed}[innertopmargin=0pt]
        \begin{mathpar}
            \inferrule[E-Owned]
            { \Delta(x) = \owned~C }
            { \Delta \vdash x : \owned~C \dashv \Delta }
            \and
            \inferrule[E-Shared]
            { \Delta(x) = \shared~C }
            { \Delta \vdash x : \shared~C \dashv \Delta }
            \and
            \inferrule[E-UniqueOwned]
            { \Delta(x) = \unique~C }
            { \Delta \vdash x : \owned~C \dashv \Delta }
            \and
            \inferrule[E-UniqueShared]
            { \Delta(x) = \unique~C }
            { {\begin{aligned}
                \Delta \vdash x : \shared~C \dashv \quad&\\[-5pt]
                \Delta[x \mapsto \shared~C]& \end{aligned}} }
            \and
            \inferrule[E-UniqueUnique]
            { \Delta(x) = \unique~C }
            { {\begin{aligned}
                \Delta \vdash x : \unique~C \dashv\quad& \\[-5pt]
                \Delta[x \mapsto \bot~C]& \end{aligned}} }
            \and
            \inferrule[E-UniqueAlias]
            { \Delta(x) = \unique~C }
            { {\begin{aligned}
                \Delta \vdash x : \unique(p.f)~C \dashv \quad& \\[-5pt]
                \Delta[x \mapsto \alias(p.f)~C]& \end{aligned}} }
            \and
            \inferrule[E-Alias]
            { \Delta(x) = \alias(p)~C \\\\ \Delta \vdash p : \alpha_e~C \dashv \Delta' }
            { \Delta \vdash x : \alpha_e~C \dashv \Delta' }
            \and
            \inferrule[E-FieldOwned]
            { \Delta \vdash p : \owned~C \dashv \Delta \\\\ \ftype(C)(f) = \unique~C' }
            { \Delta \vdash p.f : \owned~C' \dashv \Delta }
            \and
            \inferrule[E-FieldShared]
            { \Delta \vdash p : C \\\\ \ftype(C)(f) = \shared~C' }
            { \Delta \vdash p.f : \shared~C' \dashv \Delta }
            \and
            \inferrule[E-Sub]
            { \Delta \vdash e : \alpha_e~C \dashv \Delta' \\\\ C <: D }
            { \Delta \vdash e : \alpha_e~D \dashv \Delta'}
            \and
            \inferrule[E-Null]
            { }
            { \Delta \vdash \knull : \alpha_e~C \dashv \Delta }
        \end{mathpar}
    \end{mdframed}
    \caption{Expression usage typing rules}
    \label{fig:expression-usage-rules}
\end{figure}

\subsubsection{Isolation}\label{sec:isolation}

$\Delta * p \dashv \Delta'$ denotes that $p$ is \emph{isolated} from $\Delta'$ -- all references to $p$ contained in $\Delta$ are removed in $\Delta'$. $\Delta'$ represents a state where $p$ is assigned a new value, thus all aliases to $p$ in $\Delta$ should be removed in $\Delta'$. Moreover, if $p$ represented a unique value and a variable $x$ was aliased to $p$, $x$ contains a unique value after $p$ is overwritten. Thus destructive reads are accomplished by first introducing an alias to $p$, and then overwriting $p$ with a different value, such as $\knull$.


Given an environment $\Delta$, we define its \textit{reference graph} to be a (directed) graph whose nodes are syntactic paths. An edge $x \to p$ exists iff $\Delta$ contains an annotation $x : \alias(p.\cdots)$.

Intuitively, the origin of an edge in the alias graph identifies a variable whose annotation requires updating when its target is mutated. Unlike the alias graph defined in \S\ref{sec:aliasing}, this graph is not symmetric or transitive.

For example, in the \ttt{pop} method, after line \ref{ln:pop-value-assign} we have the annotation $\ttt{value} : \alias(\ttt{this.root.value})$. Thus the reference graph contains the edge $\ttt{value} \to \allowbreak \ttt{this.root}$. If the value of \ttt{this.root} is changed, we must determine a new annotation for \ttt{value}.

Formal rules are given in appendix \ref{sec:isolation-rules}. The rules deal with three main cases:
\begin{enumerate}
    \item No node in either the reference graph or the alias graph $p$ is connected to $p$. $\Delta$ is unchanged, except to remove $p$ from $\dom(\Delta)$. (See the \textsc{I-Remove*} rules.)
    
    This case is applied during validation of line \ref{ln:pop-value-assign} in \Cref{fig:stack-example} to remove the initial annotation $\ttt{value} : \bot$.
    
    \item For some variable $x$ (distinct from $p$), $x$ is aliased with $p$. In this case, all paths rooted in $p$ may be replaced by paths rooted in $x$. (See the \textsc{I-Replace*} rules.)

    \item $p$ is disconnected in the alias graph, but $p$ is the target of an edge in the reference graph. In this case, we can isolate the subfield that induces this edge (such as $p.f$) before isolating $p$. (See the \textsc{I-Elim*} rules.)
    
    This case is applied during validation of line \ref{ln:pop-next-update} in \Cref{fig:stack-example} when isolating \ttt{this.root}. The annotation of \ttt{value} is updated from
    $\alias(\ttt{this.root.value})$ to $\unique$.
\end{enumerate}

\subsubsection{Framing}\label{sec:framing}

When a reference is passed to a method, any field of the referenced object can be modified. The \emph{frame} of a method call contains all such fields. Any aliases to fields in the method's frame must be invalidated after the method call. 


$\Delta \star \m{e} \dashv \Delta'$ denotes that all variables that are connected (in the reference graph of $\Delta$) to any path in the list $\m{e}$ are marked in $\Delta'$ as either $\shared$, in cases where alias tracking is unnecessary, or $\bot$, in cases where aliases must be tracked. Formal rules are given in appendix \ref{sec:framing-rules}.

\subsubsection{Statement typing}\label{sub-sec:statement-typing}

Statements are typed by the judgment $\Delta \vdash s \dashv \Delta'$. Selected rules are shown in \Cref{fig:stmt-typing}. See appendix \ref{sec:stmt-typing-rules} for a complete listing.

An assignment such as $x \kassign p$ adds an edge $x \to p$ to the alias graph. Note that as a special case, $x \kassign \knull$ adds the annotation $x : \unique$ (see definition of $\aliasfn$ in \Cref{ref:auxiliary-fns}).

When $f$ is a $\unique$ field and $e$ is a path, $x.f \kassign e$ adds an edge $e \to x.f$ to the alias graph by using the annotation $\unique(x.f)$ when validating $e$. Note that $x.f$ is isolated \emph{before} typing $e$, which allows patterns such as lines \ref{ln:pop-next-assign}-\ref{ln:pop-next-update} in \Cref{fig:stack-example}, where \ttt{this.root} is updated to point to \ttt{this.root.next} without a destructive read. Since the field \ttt{this.root} is $\unique$, the variable aliased to \ttt{this{\allowbreak}.root.next} may overwrite \ttt{this.root} without violating uniqueness.

Method calls are validated by validating each argument according to the corresponding parameter definition. Note that after typing an argument as $\unique$, all paths rooted in the argument are inaccessible. Therefore reachable aliasing cannot exist between parameters annotated with $\unique$. However, aliasing is not prohibited simply by typing arguments passed to parameters annotated $\owned$, since borrowed values are not consumed. Therefore we explicitly prohibit reachable aliasing between $\owned$ parameters. We also invalidate aliases to fields in the frame of the method call, as discussed in \S\ref{sec:framing}.

\begin{figure}
    \begin{mdframed}[innertopmargin=0pt]
        \begin{mathpar}
            \inferrule[S-AssignVar]
            {
                \Delta(x) = \alpha~C \\
                \Delta \vdash e : C \\
                \Delta \nvdash x \equiv e \\
                \Delta * x \dashv \Delta'
            }
            {
                \Delta \vdash x \kassign e\semi \dashv x : \aliasfn(e)~C, \Delta'
            }
            \and
            \inferrule[S-Call]
            {
                \alpha_0~C_0, \cdots, \alpha_n~C_n \to \alpha~C = \mtype(C_0)(m) \\
                \Delta \vdash e_0 : \alpha_0~C_0, \cdots, e_n : \alpha_n~C_n \dashv \Delta' \\
                \Delta'(x) = \alpha'~B \\ B <: C \\
                \Delta' \star e_0, \cdots, e_n \dashv \Delta'' \\
                \Delta'' * x \dashv \Delta''' \\
                \forall 0 \le i < j \le n : \big[ \alpha_i = \owned \implies \Delta' \nvdash e_i \approxeq e_j \big]
            }
            {
                \Delta \vdash x \kassign e_0.m(e_1, \cdots, e_n)\semi \dashv x : \alpha~B, \Delta'''
            }
            \and
            \inferrule[S-AssignUnique]
            {
                \Delta \vdash x : C \\
                \Delta * x.f \dashv \Delta' \\
                \ftype(C)(f) = \unique~C' \\
                \Delta \nvdash x.f \approxeq e \\
                \Delta' \vdash e : \unique(x.f)~C' \dashv \Delta''
            }
            {
                \Delta \vdash x.f \kassign e\semi \dashv \Delta''
            }
            \and
            \inferrule[S-Conditional]
            {
                \Delta \vdash e_1 : \ttt{Object} \\
                \Delta \vdash e_2 : \ttt{Object} \\\\
                \Delta \vdash s_1 \dashv \Delta_1 \\
                \Delta \vdash s_2 \dashv \Delta_2
            }
            {
                \Delta \vdash \kif~(e_1 \keq e_2)~s_1~\kelse~s_2 \dashv \unify(\Delta; \Delta_1; \Delta_2)
            }
        \end{mathpar}
    \end{mdframed}
    \caption{Statement typing rules}\label{fig:stmt-typing}
\end{figure}

\subsubsection{Unification}\label{sec:unification}

When an $\kif$ statement is validated, each branch must be validated, which produces two separate typing environments, $\Delta_1$ and $\Delta_2$. These environments are \textit{unified} by finding an annotation that can be used in both environments for each variable in the outer environment $\Delta$. Formal rules are given in appendix \ref{sec:unification-rules}.

If some variable $x$ has the same annotation in both branches, then this common annotation may be used. If $x$ refers to a $\shared$ value in both $\Delta_1$ and $\Delta_2$, then it may be annotated with $\shared$. If a variable is not referenced in starting $\Delta$, the variable may be isolated from $\Delta_1$ and $\Delta_2$. This removes all aliasing to variables defined within a particular branch.

If $x$ is annotated $\alias(p_1)$ in $\Delta_1$ and $\alias(p_2)$ in $\Delta_2$, and there is a path $p$ where $\Delta_1 \vdash p_1 \equiv p$ and $\Delta_2 \vdash p_2 \equiv p$, then $x$ may be annotated with $\alias(p)$ after unification. Otherwise, if none of these cases apply, the variable is annotated with $\bot$, thus making its value inaccessible.

Unification occurs in the \ttt{pop} method in \Cref{fig:stack-example} prior to the \ttt{return} statement. In the first branch, \ttt{value} is $\unique$ since it is $\knull$. In the second branch, \ttt{value} is first aliased with \ttt{this.root.value}, but then becomes $\unique$ after line \ref{ln:pop-next-update}. Thus \ttt{value} is $\unique$ in both branches, and we can use \ttt{value} in the \ttt{return} statement, which requires a $\unique$ value.

Note that more specific unification procedures could be developed. For example, we could implement conditional aliasing annotations, or validate every possible execution path independently (and thus eliminate unification entirely). However, we feel that this conservative unification algorithm is more usable since it allows clear error messages when attempting to access variables that are inaccessible due to unification.

\subsubsection{Program typing}

A valid program is defined in appendix \ref{sec:program-typing-rules}, along with rules for typing method and class declarations.

\subsection{Extended example}

The \ttt{dequeue} method in \Cref{fig:dequeue-example} allows the stack to be used as a FIFO queue. The recursive traversal of the linked list is handled by the \ttt{dequeueHelper} method. Note that the entire list is traversed, and the tail modified, using only a single destructive read.

This is enabled by borrowing the $\unique$ value \ttt{this.root}, and in turn each \ttt{next} node. The value of an $\owned$ parameter is guaranteed to be unique, but its value may not be consumed or placed on the heap. Thus \ttt{dequeueHelper} guarantees that no additional aliases to \ttt{n} will be introduced. However, the contents of an $\owned$ value may be modified, which allows the tail \ttt{Node} to be removed.


Also note that the $\unique$ value \ttt{n.next.value} is read without an explicit destructive read at line \ref{ln:dequeue-value-assign}. Instead, it is known to be unique since its container (\ttt{n.next}) is isolated at line \ref{ln:dequeue-next-assign}.

In a survey of related work, \citet{DBLP:conf/pldi/MilanoTM22} found that ownership systems often require explicit destructive reads at each step when traversing and modifying a linked list as in this example. 
However, our isolation technique, combined with local aliased values and borrowing, eliminates this requirement while allowing common code patterns and requiring few annotations.

\begin{figure}
\begin{mdframed}[innertopmargin=7pt,innerleftmargin=15pt,innerrightmargin=0pt,innerbottommargin=7pt]
{\footnotesize
\begin{lstlisting}[aboveskip=0pt,belowskip=0pt]
unique Object dequeue(owned Stack this) {
  Node r = this.root;
  Object value;
  if (r == null || r.next == null) {
    value = this.pop();
  } else {
    value = dequeueHelper(r);
  }
  return value;
}

unique Object dequeueHelper(owned Stack this,
                                     owned Node n) {
  Object value;
  if (n.next.next == null) {
    value = n.next.value; <@\label{ln:dequeue-value-assign}@>
    n.next = null; <@\label{ln:dequeue-next-assign}@>
  } else {
    value = dequeueHelper(n.next);
  }
  return value;
}
\end{lstlisting}}
\end{mdframed}
    \caption{Example: a \ttt{dequeue} method for the \ttt{Stack} class from \Cref{fig:stack-example}}
    \label{fig:dequeue-example}
\end{figure}

\section{Future Work}\label{sec:futurework}

While the design decisions were guided by making this system usable by developers, we would need to implement the system and evaluate it in larger examples with users. A comparison of the effort in different examples with the alternatives in the related work would also be interesting, to confirm whether our invariant helps developers to annotate their code.
Additionally, our current approach includes only a core set of Java features that we would like to extend to include \ttt{while} loops.

One of the motivations of this work was to introduce enough information to reason about mutability to support Liquid Types in a mutable context. Flux~\cite{Lehmann2022FluxLT} took the first steps in this area, by using Rust's ownership type in combination with a Liquid Type System. Our proposal targets the Java language instead, and serves as the basis to extend LiquidJava~\cite{liquidjava} to better model aliasing and uniqueness combined with refinements. Because Liquid Types supports a logic-based version of symbolic execution, the information from refinements could be used in the unification to have a more precise alias tracking, instead of the conservative invalidation we took instead.

\section{Conclusion}

We have described Latte, a simple type system for uniqueness and aliasing for Java, which prioritizes usability and low development overhead. Our vision is that more complex type systems may utilize the uniqueness and aliasing information determined by Latte.
Latte enforces (and requires consideration of) simple invariants of values on the heap, imposes a low annotation burden, and requires no changes to existing Java semantics.

Our simple uniqueness invariant indicates that a unique object is stored at most once on the heap. In addition, all usable references to a unique object from the local environment are precisely inferred.
The developer only needs to annotate field declarations and the parameters and return types of method declarations, using one of $\unique$, $\owned$ or $\shared$.

While it may lack the expressive power of related approaches, we hope that Latte provides a lower barrier to entry for existing Java developers, thus enhancing the appeal of automated verification tools built on Latte. Further evaluation of its usability, along with the development of such verification tools, is required to validate this goal.

\section{Acknowledgments}


This work was supported by the National Science Foundation under Grant No. CCF-1901033 and by the Algorand Centres of Excellence programme managed by Algorand Foundation. 
The work is also co-financed by a Dual Degree Ph.D. Scholarships awarded by the Portuguese Foundation for Science and Technology through the Carnegie Mellon Portugal Program, through the RAP project (EXPL/CCI-COM/1306/2021) and through the LASIGE unit (UIDB/00408/2020 and UIDP/00408/2020).
Any opinions, findings, and conclusions or recommendations expressed in this material are those of the author(s) and do not necessarily reflect the views of the NSF or the Algorand Foundation.

\bibliographystyle{ACM-Reference-Format}
\bibliography{bib-aliasing}

\clearpage
\appendix

\section{Typing rules}

\subsection{Reference typing}\label{sec:reference-typing}

$\Delta \vdash p : C$ denotes that the path $p$ is accessible and contains either $\knull$ or an reference to an instance of class $C$.

\begin{mathpar}
    \inferrule[T-Var]
    { \alpha \ne \bot }
    { \Delta, x : \alpha~C \vdash x : C }
    \and
    \inferrule[T-Field]
    { \Delta \vdash p : C \\\\ \ftype(C)(f) = \alpha~C' }
    { \Delta \vdash p.f : C' }
\end{mathpar}

\subsection{Subtyping}\label{sec:subtyping}

$C <: D$ denotes that $C$ extends a parent class $D$. It is a reflexive and transitive relation.

\begin{mathpar}
    \inferrule[C-Id]
        {  }
        { C <: C }
    \and
    \inferrule[C-Transitive]
        { C <: D \\ D <: E }
        { C <: E}
    \and
    \inferrule[C-Def]
        { \kclass C \kextends D ~ \block{ \m{F} ~ K ~ \m{M} }}
        { C <: D }
\end{mathpar}

\subsection{Auxiliary functions}\label{ref:auxiliary-fns}

$\fields(C)$ denotes the list of field declarations in $C$. $\ftype(C)(f)$ denotes the annotation and class specified in the declaration of field $f$ in class $C$. $\mtype(C)(m)$ denotes the signature of method $m$ in class $C$. $\aliasfn(e)$ is a helper function which denotes the annotation $\alias(e)$ when $e$ is a path, and $\unique$ when $e$ is $\knull$.

\begin{mathpar}
    \inferrule[F-Obj]
        {  }
        { \fields (\ttt{Object}) = \cdot}
    \and
    \inferrule[F-Decl]
        { 
        \kclass C \kextends D ~ \block{ \m{F} ~ K ~ \m{M} }\\
        \m{F} = \m{\alpha_{f} ~C~ f}\semi\\
        \fields(D) = \m{\alpha_{f'} ~D~ g}
        }
        { \fields (C) = \m{\alpha_{f'} ~D~ g}~~\m{\alpha_{f} ~C~ f}}
    \and
    \inferrule[F-Type]
    { \alpha_f ~B~f \in \fields (C)}
    { \ftype (C) (f) = \alpha_f ~B}
    \and
    \inferrule[M-Class]
        { 
        \kclass C \kextends D ~ \block{ \m{F} ~ K ~ \m{M} }\\\\ 
        \tau ~ m(\alpha_p~\kthis, ~ \m{\alpha_p ~ B ~ x}) ~ \block{ \m{s} } \in \m{M}
        }
        { \mtype (C) (m) =  \m{\alpha_p~B} \to \tau }
    \and
    \inferrule[M-Super]
    { 
        \kclass C \kextends D ~ \block{ \m{F} ~ K ~ \m{M} }\\ 
        \mtype (D)(m) =  \m{\alpha_p~B} \to \tau
    }
    { \mtype (C) (m) =  \m{\alpha_p~B} \to \tau} \and
    \aliasfn(e) := \begin{cases}
        \unique &\text{if $e = \knull$} \\
        \alias(e) &\text{otherwise}
    \end{cases}
\end{mathpar}

\subsection{Aliasing}\label{sec:aliasing-rules}

$\Delta \vdash p \equiv p'$ denotes that a path $p$ refers to the same value as $p'$.

\begin{mathpar}
    \inferrule[A-Refl]
        { }
        { \Delta \vdash p \equiv p }
    \and
    \inferrule[A-Var]
        { \Delta(x) = \alias(p)~C }
        { \Delta \vdash x \equiv p }
    \and
    \inferrule[A-Cong]
        { \Delta \vdash p_1 \equiv p_2 }
        { \Delta \vdash p_1.f \equiv p_2.f }
    \and
    \inferrule[A-Symm]
        { \Delta \vdash p_1 \equiv p_2 }
        { \Delta \vdash p_2 \equiv p_1 }
    \and
    \inferrule[A-Trans]
        { \Delta \vdash p_1 \equiv p_2 \\ \Delta \vdash p_2 \equiv p_3 }
        { \Delta \vdash p_1 \equiv p_3 }
\end{mathpar}

\subsection{Reachable aliasing}\label{sec:reachable-aliasing-rules}

$\Delta \vdash p \approxeq p'$ denotes that a value reachable from $p$ may be aliased with a value reachable from $p'$, given the aliasing specified in $\Delta$.

\begin{mathpar}
    \inferrule[RA-Refl]
        { }
        { \Delta \vdash p \approxeq p }
    \and
    \inferrule[RA-Symm]
        { \Delta \vdash p_1 \approxeq p_2 }
        { \Delta \vdash p_2 \approxeq p_1 }
    \and
    \inferrule[RA-Trans]
        { \Delta \vdash p_1 \approxeq p_2 \\
          \Delta \vdash p_2 \approxeq p_3 }
        { \Delta \vdash p_1 \approxeq p_3 }
    \and
    \inferrule[RA-Field]
        { \Delta \vdash p_1 \approxeq p_2 }
        { \Delta \vdash p_1.f \approxeq p_2 }
    \and
    \inferrule[RA-Var]
        { \Delta(x) = \alias(p)~C }
        { \Delta \vdash x \approxeq p }
\end{mathpar}

\subsection{Replacement}\label{sec:replacement-rules}

$p[p'' / p']$ denotes the path $p$ with the path $p'$ replaced by $p''$. For example, $x.f[y.g / x] = y.g.f$.

\begin{align*}
x[p''/p'] &:= \begin{cases}
    p'' & \text{if $x = p'$} \\
    x & \text{otherwise}
\end{cases} \\
p.f[p'' / p'] &:= \begin{cases}
    p'' &\text{if $p.f = p'$} \\
    p[p'' / p'].f &\text{otherwise}
\end{cases}
\end{align*}

$\Delta[p'' / p']$ denotes the environment $\Delta$ with all aliases that reference the path $p'$ updated to instead reference the path $p''$.

\begin{align*}
    \cdot [p'' / p'] &:= \cdot \\
    (x : \alpha~C, \Delta)[p'' / p'] &:= \left\{
        \begin{array}{l}
        x : \alias(p[p'' / p'])~C, \Delta[p'' / p'] \\
            \quad\text{if $\alpha = \alias(p)$ for some $p$} \\
        x : \alpha~C, \Delta'[p'' / p'] \\
            \quad\text{otherwise}
    \end{array} \right.
\end{align*}

\subsection{Isolation}\label{sec:isolation-rules}

$\Delta \vdash x \rightsquigarrow p$ denotes that a variable $x$ is aliased with a path rooted in $p$, i.e. $p$, $p.f$, $p.f.g$, etc. In other words, an aliased edge exists between $x$ and $p$ or a reference edge points to $p$.

\begin{mathpar}
    \inferrule[VA-Intro]
        { \Delta(x) = \alias(p') \\ \Delta \vdash p \equiv p' }
        { \Delta \vdash x \rightsquigarrow p } \and
    \inferrule[VA-FieldElim]
        { \Delta \vdash x \rightsquigarrow p.f }
        { \Delta \vdash x \rightsquigarrow p }
\end{mathpar}

$\Delta * p \dashv \Delta'$ denotes that a path $p$ is isolated from $\Delta'$, as described in \Cref{sec:isolation}.

\begin{mathpar}
    \inferrule[I-RemoveField]
        { \nexists x \in \dom(\Delta) : \Delta \vdash x \rightsquigarrow p.f }
        { \Delta * p.f \dashv \Delta }
\end{mathpar}
\begin{mathpar}
    \inferrule[I-ReplaceUnique]
        { \Delta = x : \alias(p'.f)~C, \Delta' \\
          \Delta \vdash p \equiv p' \\
          \Delta \vdash p.f : \owned~C' \dashv \Delta }
        { \Delta * p.f \dashv x : \unique~C, \Delta'[x / p.f] }
\end{mathpar}
\begin{mathpar}
    \inferrule[I-ReplaceShared]
        { \Delta = x : \alias(p'.f)~C, \Delta' \\
          \Delta \vdash p \equiv p' \\
          \Delta \vdash p.f : \shared~C' \dashv \Delta }
        { \Delta * p.f \dashv x : \shared~C, \Delta'[x / p.f] }
\end{mathpar}
\begin{mathpar}
    \inferrule[I-ReplaceInaccessible]
        { \Delta = x : \alias(p'.f)~C, \Delta' \\
          \Delta \vdash p \equiv p' }
        { \Delta * p.f \dashv x : \bot~C, \Delta'[x / p.f] }
\end{mathpar}
\begin{mathpar}
    \inferrule[I-ElimField]
        { \nexists x \in \dom(\Delta) : \Delta(x) \vdash x \equiv p.f \\
          \Delta \vdash x \rightsquigarrow p.f.g \\
          \Delta * p.f.g \dashv \Delta' \\
          \Delta' * p.f \dashv \Delta'' }
        { \Delta * p.f \dashv \Delta'' }
\end{mathpar}
\begin{mathpar}
    \inferrule[I-RemoveVar]
        { \Delta = x : \alpha~C, \Delta' \\
          \nexists y \in \dom(\Delta') : \Delta \vdash y \rightsquigarrow x }
        { \Delta * x \dashv \Delta' }
\end{mathpar}
\begin{mathpar}
    \inferrule[I-ReplaceAlias]
        { \Delta = x : \alias(p)~C, \Delta' }
        { \Delta * x \dashv \Delta'[p / x] }
\end{mathpar}
\begin{mathpar}
    \inferrule[I-ReplaceAliased]
        { \Delta = x : \alpha~C, y : \alpha'~C', \Delta' \\
          \Delta \vdash y \equiv x \\
          \alpha \in \set{ \unique, \owned, \shared, \bot } }
        { \Delta * x \dashv y : \alpha~C, \Delta'[y / x] }
\end{mathpar}
\begin{mathpar}
    \inferrule[I-ElimVar]
        { \Delta = x : \alpha~C, \Delta' \\
          \nexists y \in \dom(\Delta') : \Delta \vdash y \equiv x \\
          \Delta \vdash y \rightsquigarrow x.f \\
          \Delta * x.f \dashv \Delta' \\
          \Delta' * x \dashv \Delta'' }
        { \Delta * x \dashv \Delta'' }
\end{mathpar}

\subsection{Framing}\label{sec:framing-rules}

$\Delta \vdash \Delta \star p \dashv \Delta'$ denotes that all edges $p$ in the reference graph of $\Delta$ are inaccessible or shared in $\Delta'$. This is used to remove alias references to fields that are contained in a method call's frame since those aliases may no longer be correct if the method changes that field.

$\Delta \star \m{p} \dashv \Delta'$ repeats this process for all paths in the list $\m{p}$.

\begin{mathpar}
    \inferrule[R-Shared]
        { \alpha \in \set{ \shared, \owned, \unique, \bot } \\
          \Delta \vdash \Delta' \star p \dashv \Delta'' }
        { \Delta \vdash x : \alpha~C, \Delta' \star p \dashv x : \alpha~C, \Delta'' } \and
    \inferrule[R-SharedAlias]
        { \Delta \vdash p' : \shared~C \dashv \Delta \\
          \Delta \vdash \Delta' \star p \dashv \Delta'' }
        { \Delta \vdash x : \alias(p')~C, \Delta' \star p \dashv x : \shared~C, \Delta'' } \and
    \inferrule[R-Separate]
        { \nexists f : \Delta \vdash x \rightsquigarrow p.f \\
          \Delta \vdash \Delta' \star p \dashv \Delta'' }
        { \Delta \vdash x : \alpha~C, \Delta' \star p \dashv x : \alpha~C, \Delta'' } \and
    \inferrule[R-Inaccessible]
        { \Delta \vdash \Delta' \star p \dashv \Delta'' }
        { \Delta \vdash x : \alpha~C, \Delta' \star p \dashv x : \bot~C, \Delta'' } \and
    \inferrule[R-Empty]
        { }
        { \Delta \vdash \cdot \star p \dashv \cdot }
\end{mathpar}
\begin{align*}
    \Delta \star p_1, \cdots, p_n \dashv \Delta' \iff \Delta = \Delta_0 &\vdash \Delta_0 \star p_1 \dashv \Delta_1 \\
        &\vdots \\
        \Delta_{n-1} &\vdash \Delta_{n-1} \star p_n \dashv \Delta_n = \Delta'
\end{align*}

\subsection{Unification}\label{sec:unification-rules}

$\alpha \preceq \alpha'$ denotes that an annotation $\alpha$ may be used in place of $\alpha'$ -- this means that $\alpha$ represents fewer permissions than $\alpha'$.

\begin{mathpar}
    \inferrule[A-Id]
    { }
    { \Delta \vdash \alpha \preceq \alpha } \and
\inferrule[A-Shared]
    { }
    { \Delta \vdash \shared \preceq \unique } \and
\inferrule[A-Inacc]
    { }
    { \Delta \vdash \bot \preceq \alpha } \and
\inferrule[A-Alias]
    { \Delta \vdash p_1 \equiv p_2 }
    { \Delta \vdash \alias(p_1) \preceq \alias(p_2) } \and
\inferrule[A-SharedAlias]
    { \Delta \vdash p.f : \shared~C \dashv \Delta }
    { \Delta \vdash \shared \preceq \alias(p.f) }
\end{mathpar}

$\unify(\Delta; \Delta_1; \Delta_2)$ denotes the unification of branches $\Delta_1$ and $\Delta_2$, given a parent environment $\Delta$.

\begin{mathpar}
    \inferrule[U-Empty]
        { }
        { \unify(\cdot; \Delta_1; \Delta_2) = \cdot } \and
    \inferrule[U-Comm]
        { \unify(\Delta; \Delta_1; \Delta_2) = \Delta' }
        { \unify(\Delta; \Delta_2; \Delta_1) = \Delta' }
\end{mathpar}
\begin{mathpar}
    \inferrule[U-Join]
        {
            \unify(\Delta; \Delta_1; \Delta_2) = \Delta' \\
            \Delta_1(x) = \alpha_1~C \\
            \Delta_2(x) = \alpha_2~C \\
            \Delta' \vdash \alpha_1 \preceq \alpha_2
        }
        {
            \unify(x : \alpha~C, \Delta_1; \Delta_2) = x : \alpha_1~C, \Delta'
        }
\end{mathpar}
\begin{mathpar}
    \inferrule[U-Isolate]
        {
            y * \Delta_1 \dashv \Delta_1' \\
            \unify(\Delta; \Delta_1'; \Delta_2) = \Delta' \\
        }
        {
            \unify(\Delta; y : \alpha~C, \Delta_1; \Delta_2) = \Delta'
        }
\end{mathpar}

\subsection{Statement typing}\label{sec:stmt-typing-rules}

A statement $s$ is valid if $\Delta \vdash s \dashv \Delta'$. $\Delta'$ represents the state after $s$ is executed.

For ease of notation, we define $\Delta \vdash e_1 : \alpha_1~C_1, \cdots, e_n : \alpha_n~C_n \dashv \Delta'$ to hold iff, for some $\Delta_0, \cdots, \Delta_n$,
\begin{align*}
    \Delta = \Delta_0 &\vdash e_1 : \alpha_1~C_1 \dashv \Delta_1 \\
        &\vdots \\
        \Delta_{n-1} &\vdash e_n : \alpha_n~C_n \dashv \Delta_n = \Delta'
\end{align*}

Similarly, $\Delta \vdash s_1, \cdots, s_n \dashv \Delta'$ holds iff, for some $\Delta_0, \cdots, \Delta_n$,
$$\Delta = \Delta_0 \vdash s_1 \dashv \Delta_1, \cdots, ~\text{and}~ \Delta_{n-1} \vdash s_n \dashv \Delta_n = \Delta'.$$

\begin{mathpar}
    \inferrule[S-Block]
    { \Delta \vdash \m{s} \dashv \Delta' }
    { \Delta \vdash \block{\m{s}} \dashv \Delta' }
    \and
    \inferrule[S-Decl]
    { }
    { \Delta \vdash C~x\semi \dashv x : \bot~C, \Delta }
\end{mathpar}
\begin{mathpar}
    \inferrule[S-AssignVar]
    {
        \Delta(x) = \alpha~C \\
        \Delta \vdash e : C \\
        \Delta \nvdash x \equiv e \\
        \Delta * x \dashv \Delta'
    }
    {
        \Delta \vdash x \kassign e\semi \dashv x : \aliasfn(e)~C, \Delta'
    }
\end{mathpar}
\begin{mathpar}
    \inferrule[S-AssignShared]
    {
        \Delta \vdash x : C \\
        \Delta * x.f \dashv \Delta' \\
        \ftype(C)(f) = \shared~C' \\
        \Delta' \vdash e : \shared~C' \dashv \Delta''
    }
    {
        \Delta \vdash x.f \kassign e\semi \dashv \Delta''
    }
\end{mathpar}
\begin{mathpar}
    \inferrule[S-AssignUnique]
    {
        \Delta \vdash x : C \\
        \Delta * x.f \dashv \Delta' \\
        \ftype(C)(f) = \unique~C' \\
        \Delta \nvdash x.f \approxeq e \\
        \Delta' \vdash e : \unique(x.f)~C' \dashv \Delta''
    }
    {
        \Delta \vdash x.f \kassign e\semi \dashv \Delta''
    }
\end{mathpar}
\begin{mathpar}
    \inferrule[S-New]
    {
        \alpha_1~C_1~f_1, \cdots, \alpha_n~C_n~f_n= \mathsf{fields}(C)\\
        \Delta \vdash e_1 : \alpha_1~C_1, \cdots, e_n : \alpha_n~C_n \dashv \Delta' \\
        \Delta'(x) = \alpha~B \\ B <: C \\
        \Delta' * x \dashv \Delta'' \\\\
        \forall 0 \le i < j \le n : \big[ \alpha_i = \owned \implies \Delta' \nvdash e_i \approxeq e_j \big]
    }
    {
        \Delta \vdash x \kassign \knew C(\m{e})\semi \dashv x : \unique~B, \Delta''
    }
\end{mathpar}
\begin{mathpar}
    \inferrule[S-Call]
    {
        \alpha_0~C_0, \cdots, \alpha_n~C_n \to \alpha~C = \mtype(C_0)(m) \\
        \Delta \vdash e_0 : \alpha_0~C_0, \cdots, e_n : \alpha_n~C_n \dashv \Delta' \\
        \Delta'(x) = \alpha'~B \\ B <: C \\
        \Delta' \star e_0, \cdots, e_n \dashv \Delta'' \\
        \Delta'' * x \dashv \Delta''' \\
        \forall 0 \le i < j \le n : \big[ \alpha_i = \owned \implies \Delta' \nvdash e_i \approxeq e_j \big]
    }
    {
        \Delta \vdash x \kassign e_0.m(e_1, \cdots, e_n)\semi \dashv x : \alpha~B, \Delta'''
    }
\end{mathpar}
\begin{mathpar}
    \inferrule[S-CallVoid]
    {
        \alpha_0~C_0, \cdots, \alpha_n~C_n \to \tau = \mtype(C_0)(m) \\
        \Delta \vdash e_0 : \alpha_0~C_0, \cdots, e_n : \alpha_n~C_n \dashv \Delta' \\
        \Delta' \star e_0, \cdots, e_n \dashv \Delta'' \\
        \forall 0 \le i < j \le n : \big[ \alpha_i = \owned \implies \Delta' \nvdash e_i \approxeq e_j \big]
    }
    {
        \Delta \vdash e_0.m(e_1, \cdots, e_n)\semi \dashv \Delta''
    }
\end{mathpar}
\begin{mathpar}
    \inferrule[S-Conditional]
    {
        \Delta \vdash e_1 : \ttt{Object} \\
        \Delta \vdash e_2 : \ttt{Object} \\
        \Delta \vdash s_1 \dashv \Delta_1 \\
        \Delta \vdash s_2 \dashv \Delta_2 \\
        \unify(\Delta; \Delta_1; \Delta_2) = \Delta'
    }
    {
        \Delta \vdash \kif~(e_1~\ttt{==}~e_2)~s_1~\kelse~s_2 \dashv \Delta'
    }
\end{mathpar}

\subsection{Program typing}\label{sec:program-typing-rules}

A program $P = \m{CL}$ is valid if $CL~\ttt{ok}$ for all class declarations $CL$.

\begin{mathpar}
    \inferrule[T-Method]
    { 
        \kthis : \alpha_{p0} ~ C,~ \m{x : \alpha_p~ C} \vdash ~ s ~ \dashv \Delta'\\\\
        \Delta' \vdash e : \alpha~C' \dashv \Delta''\\\\
        \kclass C \kextends D ~ \block{...}\\
        {\begin{aligned}
            &\text{if $\mtype(D)(m) = \m{\alpha_{p}'~D} \to \tau_d$, then} \\[-5pt]
            &\quad \m{\alpha_p ~ C} = \m{\alpha_{p}'~D} \text{ and } \alpha~C' = \tau_d
        \end{aligned}}
    }
    {  
        \alpha~C' ~ m(\alpha_{p0}~\kthis, ~ \m{\alpha_p ~ C ~ x}) ~ \block{ \m{s}~ \kreturn~e\semi } \\  \ttt{ ok in } C
    }
    \and
    \inferrule[T-VoidMethod]
    { 
        \kthis : \alpha_{p0} ~ C,~ \m{x : \alpha_p~ C} \vdash ~ s ~ \dashv \Delta'\\\\
        \kclass C \kextends D ~ \block{\cdots}\\
        \ttt{if} ~ \mtype(D)(m) = \m{\alpha_{p}'~D} \to \tau_d, \ttt{, then } \\
        \m{\alpha_p ~ C} = \m{\alpha_{p}'~D} \ttt{ and } \ttt{void} = \tau_d
    }
    {  
        \ttt{void} ~ m(\alpha_{p0}~\kthis, ~ \m{\alpha_p ~ C ~ x}) ~ \block{ \m{s} } \ttt{ ok in } C
    }
    \and
    \inferrule[T-Class]
    { 
        K = C ( \m{\alpha_{f'} ~D~ g}, ~ \m{\alpha_{f} ~C~ f})\{ ~ \ksuper(\m{g}); ~ this.\m{f}=\m{f};\}\\
        \m{F} = \m{\alpha_{ f} ~C~ f}\semi\\
        \fields(D) = \m{\alpha_{f'} ~D~ g}\\
        \m{M} \ttt{ ok in C}
    }
    {  \kclass C \kextends D ~ \block{ \m{F} ~ K ~ \m{M} }\ttt{ ok}}
\end{mathpar}

\end{document}